# Accuracy of translation on ribosome could be provided by a resonance of intramolecular oscillations in tRNA molecules

Denis Semyonov.

*E-mail address*: dasem@mail.ru

*X-ray data indicate that complexes of ribosomes with cognate and near cognate tRNAs are very similar structurally, and this was the ground for a suggestion that the ribosome discriminates correct codon-anticodon pair because of its higher stability. Here an alternative explanation of kinetic proofreading is suggested, and intramolecular oscillations in tRNAs play a keystone role in it. Resonance of the oscillations allows the cognate codon-anticodon pair to be conserved due to fast energy transfer to other part of the tRNA molecule. This mechanism can potentially discriminate correct pair from an incorrect one even if they have similar stabilities.*

The results of the brilliant studies by Demeshkina et al. unambiguously showing that non-canonical GU base pair is a mimic of the canonical GC pair under conditions of translation [1, 2] is essentially significant. These findings give a good explanation for existence of near cognate codon-anticodon pairs. For all that, a statement on an unstable nature of GU pairs in their tautomeric forms seems questionable.

Of course, an expression "rare tautomeric form" became so usual [3] that instability of these pairs is implied as obvious. However, the fact that such pair could be obtained as the result of crystallization could be interpreted as an indication for its thermodynamic stability. Experimental data of other authors [4-6] show a stable nature of GU-enol pairs under conditions of replication and translation. Thermodynamically stable nature of these pairs is supported also by their existence is structures of model duplexes. So, in the Database of RNA Base-pair Structures [7] pairs that could be classified as those composed by tautomeric forms of the nucleosides are present in the structures UI_2, GP_12, GP_4, GP_5, GU_143, GU_170, GU_2, GU_171, GU_318, GU_319, GU_94 and GU_98. In this database one reference number, as a rule, corresponds to several one-type structures found in different samples (crystals of model duplexes or small RNA molecules such as tRNA).

Moreover, two-dimensional 1H15N NMR data with samples containing GU pairs [8-12] indicate that GU-enol pairs are stable beyond the crystal structures. All these spectra contain correlated signals interpreted by the authors as G and U. To my opinion, these data correspond to GU-enol pairs existing in an equilibrium with UG-enol ones (Figure 1). The presence of several pairs of such correlated signals in the spectra of tRNA could be assigned to enol pair guanine-pseudouracil, which is in a complete agreement with the respective X-ray data [7].

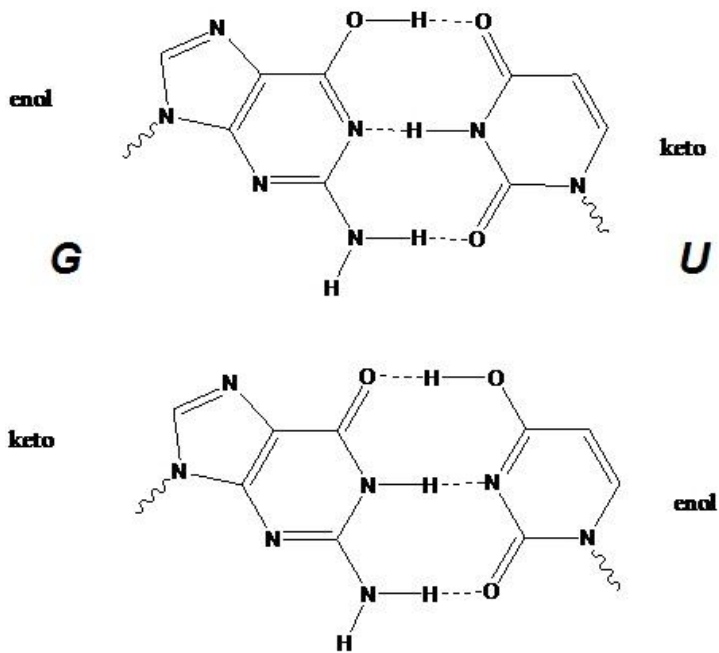

Figure 1. Two alternative variant of GU-enol structure.

So, one can suggest that GU-enol pair in [1, 2] was in fact thermodynamically stable, which is in a good agreement with a possibility of confusion of GU-enol and GC-pairs itself. Authors [1,2] use thermodynamic evaluations to justify repulsion of GU-enol pair and assume that this is essential for the possibility of discrimination cognate and near cognate tRNAs. It is generally accepted that this discrimination requires energy consumption and occurs after the GTP hydrolysis (the kinetic proofreading). As far as I understood, estimates [1, 2] do not account for the kinetic character of the process making possible discrimination of cognate and near cognate tRNA. Thereby, authors [1, 2] suggestions seem to go beyond the data justified completely by experiments.

Demeshkina et al. have demonstrated in a conclusive way that:

1. GU pair if occurs in the first or the second position of the codon-anticodon duplex, mimics the GC pair geometry. It seems probable that in this case the GU pair is in the tautomer form.

2. The mentioned GU pair is enough stable to be studied by X-ray.

3. Domain closure takes place at the recognition of cognate and some near cognate tRNA, but this rearrangement is not a key moment of the proofreading. Domain closure is necessary but not sufficient condition for the aminoacyl- tRNA accommodation

The mentioned results do not contradict my suggestion on a stable nature of GU-enol pair. I suggest that differences between GC pairs and GU-enol pairs become significant only upon excitation of the system as the result of GTP hydrolysis. Analogous suggestion concerns differences between pairs AU and GU-enol. Below, I present a proofreading mechanism that do not require thermodynamic instability of GU-enol pairs. The suggested scheme of the kinetic proofreading is supported by several peculiarities of the tRNA structure. This scheme is based on a resonance that could be observed in a system of two mutually bound linear oscillators. When parameters of the oscillators are close to each other and the bonding force is weak, almost complete energy transfer from one oscillator to another and vice versa is possible [13]. Roles of oscillators in our case are played by the anticodon and T hairpins of the tRNA (Figure 2).

Before codon-anticodon base pairing occur, these oscillators are bound to each other but could not be in the resonance since the T hairpin is fixed by the interaction with the D-loop while the anticodon hairpin is free. After formation of the codon-anticodon duplex, parameters of the oscillators become similar. They could be imagined as springs with the fixed ends.

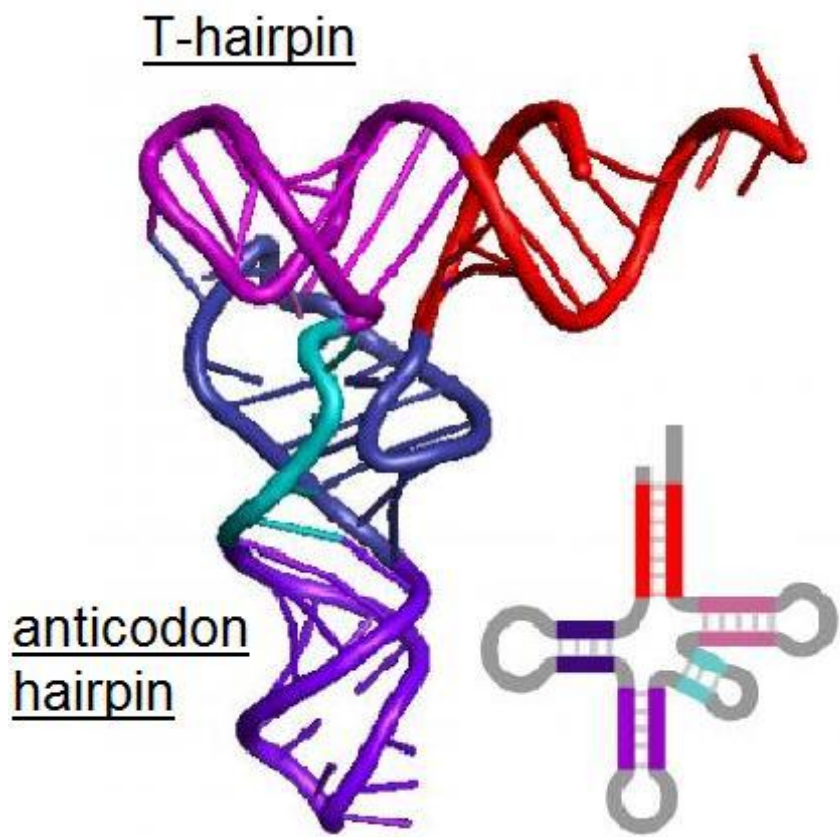

Figure 2. Spatial structure of tRNA. Anticodon and T- hairpins are violet and pink, respectively. The resonance of oscillations of these structures could provide fidelity of translation.

It is well known that spatial structures of these tRNA fragments are very similar to each other [14]: both consist of the same number of nucleotides and their loops have peculiar kink. What else, besides the structures similarity, could be an indication that anticodon and T- hairpins can interact as oscillators tuned to a resonance? The pre-requisite for the resonance to occur is similarity of physical properties of structural elements regarded as potential oscillators. Properties of anticodon and T- hairpins can be compared through a number of genomes, and one can find similarity between these tRNA fragments in almost all cases, which in turn denotes mechanism obvious with all genomes. In classical mechanics oscillator has two parameters, either the length and the mass (for a Galilleo's one) or the rigidity of the spring and the mass of the load (for a spring oscillator). In our case both the size and the masses of oscillating moieties are very similar, and parameter, which could significantly affect the oscillations is rigidity of the springs. The latter is dependent both on the number of hydrogen bonds in the hairpins and on the stacking between the nucleotides. In further discussion I take into account only the number of hydrogen bonds.

I have found two phenomena that I consider as evidence for my suggestion on resonance in oscillations of anticodon and T- hairpins.

The first one concerns tRNA(Lys) (anticodons UUU и CUU) and tRNA(Arg) (anticodons CCU and UCU). Analysis of secondary structures of the respective tRNAs shows that replacement of CUU anticodon with the UUU one is accompanied with a single replacement in the T stem that provides an additional hydrogen bond; this is seen in almost all genomic sequences from the tRNA structure database. In many cases the respective replacement in the only one and the rest part of the tRNA structure remains the same. An advantage of comparison of these tRNAs within one genome is high extent of homology of the compared tRNA sequences, which makes it possible to notice a regularity of a difference in a single hydrogen bond. So, small alteration in the anticodon results in the alteration in the T stem, and the latter compensates the former by the respective changes in number of hydrogen bonds formed.

The second phenomenon is more general and concerns all codons, but is the most pronounced with GC-rich anticodons Pro, Ala, Gly, Arg.

Analysis similar to that described above shows that with GC-rich anticodons (Pro, Ala, Gly, Arg) the T-stem is stabilized by greater number of hydrogen bonds than the anticodon stem. It is known that the interaction between the T and the anticodon hairpins involve two nucleotides from each tRNA region, while the anticodon interacts with codon via three nucleotides. Therefore, with the mentioned tRNAs, the number of hydrogen bonds in the codon-anticodon duplex is significantly greater than that in the connection between the T and the D hairpins.

In frames of the suggested hypothesis, parameters of the oscillators should be similar to each other to reach the resonance, and one of these parameters is rigidity of the "springs" that is defined by the number of hydrogen bonds. In the case of GC-rich anticodons, the excessive rigidity of the codon-anticodon interactions is equilibrated with less number of hydrogen bonds in the anticodon stem.

With GC-poor anticodons, differences between contacts codon-anticodon and contacts between T and D stems is not large, and there is no need to correct rigidity of the anticodon stem. Indeed, with AU-rich anticodons the mentioned difference in number of hydrogen bonds does not exist or even changes its sign to negative. Thus, mechanism of compensation of the oscillator's rigidity is general with both AU-rich and GC-rich anticodons.

So, I showed that there are two fragments in tRNA that can function as oscillators, and below I suggest the mode of their work in the course of kinetic proofreading.

GTP hydrolysis occurs close to the acceptor stem of tRNA, which is accompanied with movement of the tRNA with respect to the ribosome [doi: **10.2210/rcsb_pdb/mom_2006_9**] Since the energy is liberated near the acceptor stem, the L-shaped form of the tRNA molecule itself makes the acceptor stem the lever.

To describe the steps of the kinetic proofreading, I use a hypothetic but probable assumption that the tRNA movement leads to rotation of anticodon stem and loop, fixed at the mRNA, which in turn results in exciting oscillations in the anticodon hairpin. Thus, GTP hydrolysis leads to exciting oscillations in the anticodon hairpin. I suggest the existence of just such type of torsional oscillations. With near cognate tRNAs, the discussed rotation or the subsequent oscillations result in breakage of hydrogen bonds in the codon-anticodon duplex. This is the rejection stage in the course of kinetic proofreading.

With cognate tRNA, the hydrogen bonds do not break, and the energy of anticodon hairpin oscillations gradually transfers to that of the T hairpin, which results in disruption of the bonds between the T and the D hairpins. The disruption can be realized because the number of hydrogen bonds between the hairpins is less than that in the correct codon-anticodon duplex, and this should lead to large scale rearrangements in the tRNA structure, which according to my hypothesis is the main event of the accommodation of aminoacyl-tRNA to the A site [15, 16]. Thus, the selectivity of proofreading in this model is based on the resonance of oscillation of two tRNA hairpins, which implicates in the explanation of the selectivity those tRNA regions that generally are not considered to be involved.

Noteworthy, the closer are parameters of the respective oscillators, the more effective is energy transfer between them. If a near cognate codon could form a duplex with the anticodon, it does not necessarily imply that parameters of the oscillators became close to each other. Discordance of the oscillators increases the time of energy transfer between them and therefore, increases probability of the breakage of hydrogen bonds in the codon-anticodon duplex. As oscillator parameters, one can consider the geometry of the duplex (e.g., in cases when GU pair is formed instead of AU one) or the stability of the pair (in the case of GU-enol pair instead of the GC one, which according to results [1, 2] have very similar spatial structures).

All the above mentioned can be summarized as a simple scheme:

1. Binding of a cognate or near cognate tRNA to the codon.

2. GTP hydrolysis near tRNA acceptor stem leading to twisting of the anticodon hairpin , and movement of the stem in the ribosome.

3. Oscillations of anticodon hairpin of cognate and near cognate tRNA differ from each other.

4. With cognate tRNA, the parameters of the anticodon hairpin oscillations are similar to those of the T hairpin. Resonance results in rapid damping of the oscillations of the anticodon hairpin and excitation of oscillations of the T hairpin.

5. Excitation of oscillations in the T hairpin leads to disruption of hydrogen bonds between the T and the D loops, which is the key event of the accommodation step.

6. With near cognate tRNA, oscillations have parameters unfavorable for resonance. Oscillations of the T hairpin do not excite and accommodation does not occur. Simultaneously, oscillations of the anticodon hairpin do not damp, which promotes disruption of hydrogen bonds between the codon and the anticodon.

7. Discrimination of a GU-pair in near cognate case can occur even with thermodynamically stable GU pair, if parameters of oscillations with this pair significantly differ from those in the cognate case.

Resonance in the suggested scheme allows rapid single dimensional transfer of excess energy at definite degree of freedom without damage of the codon-anticodon duplex. If the suggested hypothesis is valid, alterations in the structure of the T hairpin could break the resonance, enlarge time of oscillations of the anticodon hairpin and thereby result in reject of a cognate aminoacyl-tRNA. Examination of the mentioned relationship could be regarded as examination of the suggested scheme.

In frames of the suggested mechanism, kinetic proofreading should be considered as a dynamic process, and the essential things are parameters of oscillations of the anticodon hairpin but not the energy of interactions between codon and anticodon. It seems hardly probable that the suggested excited state of tRNA could be crystallized for X-ray analysis. Besides, I hope that the suggested scheme enables discrimination of duplexes similar in their energy but differing in their dynamics. E.g., one can expect similar thermodynamic properties of AU and GU-enol pairs, but their differences in geometry should lead to differences in the oscillation parameters. Consequently, these pairs should oscillate in different ways and thereby be discriminated. Since the scheme can discriminate two states with similar energies, there is no necessity to assume that GU-enol pair is thermodynamically instable.

During last decade, single molecule spectroscopy methods have been applied for direct measurements of tRNA dynamics in the ribosome [16]. These methods seem appropriate to examine the suggested hypothesis on resonance of oscillations in the tRNA structure in the course of translation.

Acknowledgements:
The author is grateful to D.M. Graifer for helpful discussion and his help in the text preparation.

References:

1. N. Demeshkina, L. Jenner, E. Westhof, M. Yusupov, G. Yusupova, A new understanding of the decoding principle on the ribosome, Nature 7393 (2012) 256-259.

2. N. Demeshkina, L. Jenner, E. Westhof, M. Yusupov, G. Yusupova, New structural insights into the decoding mechanism: Translation fidelity via a G_U pair with Watson–Crick geometry, FEBS Lett. 587 (2013) 1848-57

3. L.A. Pray, DNA replication and causes of mutation, Nature Education (2008) 1(1)

4. K. Bebenek, L.C. Pedersen, T.A. Kunkel, Replication infidelity via a mismatch with Watson–Crick geometry, PNAS 108 (2011) 1862-1867.
5. W. Wang, H.W. Hellinga, L.S. Beese, Structural evidence for the rare tautomer hypothesis of spontaneous mutagenesis. Proc Natl Acad Sci USA. 108 (2011) 17644-17648
6. F.A.P. Vendeix, F.V. Murphy IV, W.A. Cantara, G. Leszczyńska, E.M. Gustilo, B. Sproat, A. Malkiewicz and P.F. Agris, Human tRNA Lys3 UUU Is Pre-Structured by Natural Modifications for Cognate and Wobble Codon Binding through Keto–Enol Tautomerism, J. Mol. Biol. 416 (2012) 467-485.

7. BPS Database of RNA Base-pair Structures. http://bps.rutgers.edu/bps (2008).

8. J. Noeske, C. Richter, M.A. Grundl, H.R. Nasiri, H. Shcwalbe, J. Wöhnert. An intermolecular base triple as the basis of ligand specificity and affinity in the guanine- and adenine-sensing riboswitch RNAs. PNAS 102 (2005) 1372–1377

9. N.B. Ulyanov, A. Mujeeb, Z. Du, M. Tonelli, T.G. Parslow, T.L. James. NMR Structure of the full-length linear dimer of stem-loop-1 RNA in the HIV-1 dimer initiation site. J Biol Chem. 281 (2006) 16168-16177

10. A. Grishaev, L. Yao, J. Ying, A. Pardi, A. Bax. Chemical shift anisotropy of imino 15N nuclei in Watson-Crick base pairs from magic angle spinning liquid crystal NMR and nuclear spin relaxation. J. Am. Chem. Soc. 131 (2009) 9490–9491

11. J. Farjon, J. Boisbouvier, P. Schanda, A. Pardi, J.P. Simorre, B. Brutscher, Longitudinal-relaxation-enhanced NMR experiments for the study of nucleic acids in solution. J Am Chem Soc. 131 (2009) 8571–8577.

12. Z.X. Hao, R. Feng, E.D. Wang, G. Zhu, 1H, 15N chemical shift assignments of the imino groups in the base pairs of Escherichia coli tRNALeu (CAG), Biomol NMR Assign, 5 (2011) 71–74

13. Iu.I. Neimark, Mathematical Models in Natural Science and Engineering. Springer (2003) pp141-159.

14. A.S. Spirin, Ribosome structure and protein biosynthesis. Benjamin/Cummings Pub. Co., Advanced Book Program, (1986) 414pp

15. S.V. Kirillov, V.B. Odinzov, Interaction of N-acetyl-phenylalanyl-tRNAPhe with 70S ribosomes of Escherichia coli. Nucleic Acids Res. 5(10) (1978) 3871–3879

16. S.C. Blanchard, H.D. Kim, R.L. Gonzalez, Jr., J.D. Puglisi, S. Chu, tRNA dynamics on the ribosome during translation, Proc Natl Acad Sci U S A. 101(35) (2004) 12893-8